\documentclass[reprint,
superscriptaddress,
amsmath,
amssymb,
prl
]{revtex4-2} 

\usepackage{graphicx}
\usepackage{physics}
\usepackage{color}
\usepackage{braket}
\usepackage{soul}
\usepackage[colorlinks=true, citecolor=blue, linkcolor=blue, urlcolor=blue]{hyperref}
\usepackage{comment}
\usepackage{xcolor}
\usepackage[normalem]{ulem}
\usepackage{xspace}
\usepackage{tabularx} 
\usepackage{booktabs}  
\usepackage{multirow}   
\usepackage{amsmath} 
\usepackage{bm}  
\usepackage{array}

\renewcommand{\qq}{\bm{q}}

\hypersetup{
     colorlinks   = true,
     linkcolor    = blue,
     citecolor    = blue,
     urlcolor     = blue,
}
\definecolor{mypink}{rgb}{0.858,0.188,0.478}

\newcommand{\QQ}{\textbf{Q}}

\makeatletter

\begin{document}
\title{Tensor Learning and Compression of N-phonon Interactions} 

\author{Yao Luo}%
\affiliation{Department of Applied Physics and Materials Science, and Department of Physics, California Institute of Technology, Pasadena, California 91125, USA}

\author{Dhruv Mangtani}%
\affiliation{Department of Applied Physics and Materials Science, and Department of Physics, California Institute of Technology, Pasadena, California 91125, USA}

\author{Shiyu Peng}%
\affiliation{Department of Applied Physics and Materials Science, and Department of Physics, California Institute of Technology, Pasadena, California 91125, USA}

\author{Jia Yao}%
\affiliation{Department of Applied Physics and Materials Science, and Department of Physics, California Institute of Technology, Pasadena, California 91125, USA}

\author{Sergei Kliavinek}%
\affiliation{Department of Applied Physics and Materials Science, and Department of Physics, California Institute of Technology, Pasadena, California 91125, USA}
\author{Marco Bernardi}
\affiliation{Department of Applied Physics and Materials Science, and Department of Physics, California Institute of Technology, Pasadena, California 91125, USA}
\email{bmarco@caltech.edu}

\begin{abstract}
\noindent 
Phonon interactions from lattice anharmonicity govern thermal properties and heat transport in \mbox{materials.} 
These interactions are described by $n$-th order interatomic force constants ($n$-IFCs), which can be viewed as high-dimensional tensors correlating the motion of $n$ atoms, or equivalently encoding $n$-phonon scattering processes in momentum space. 
Here, we introduce a tensor decomposition to efficiently compress $n$-IFCs for arbitrary order $n$. 
Using tensor learning, we find optimal low-rank approximations of $n$-IFCs by solving the resulting optimization problem. 
Our approach reveals the inherent low dimensionality of phonon-phonon interactions and allows compression of the 3 and 4-IFC tensors by factors of up to $10^3-10^4$ while retaining high accuracy in calculations of phonon scattering rates and thermal conductivity. 
Calculations of thermal conductivity using the compressed $n$-IFCs achieve a speed-up by nearly three orders of magnitude with >98\% accuracy relative to the reference uncompressed solution. These calculations include both 3- and 4-phonon scattering and are shown for a diverse range of materials (Si, HgTe, MgO, TiNiSn and \textcolor{black}{monoclinic ZrO$_2$}). 
In addition to accelerating state-of-the-art thermal transport calculations, the method shown here paves the way for modeling strongly anharmonic materials and higher-order phonon interactions. 
\end{abstract}

\maketitle

Phonon-phonon (ph-ph) interactions are crucial to understanding thermal transport, lattice dynamics, and structural phase transitions in condensed matter~\cite{Born}. 
They originate from the anharmonicity of the lattice potential and can be described by $n$-th order interatomic force constant ($n$-IFC) tensors, where $n\!\geq\!3$. 
Density function theory (DFT)~\cite{martin_2004} calculations combined with fitting algorithms provide accurate $n$-IFC tensors~\cite{alamode,compress-sensing-for-Anharmonicity,tdep-3ph, tdep-4ph}, enabling quantitative predictions of thermal transport in materials~\cite{single-mode-approximation,ph-ph-scattering-1,thermal-trans-2,thermal-trans-3,ShengBTE_2014,Donadio}.
However, the high-dimensionality of the IFC tensors obscures the underlying physics and poses significant computational challenges. 
The complexity of $n$-phonon ($n$-ph) interactions grows exponentially with order $n$, a clear example of the curse of dimensionality. 
For thermal conductivity calculations, 4-ph interactions require orders of magnitude more computational effort than 3-ph interactions, whereas 5-ph and higher-order ph-ph interactions remain inaccessible.
\\
\indent
To overcome this complexity, discovering low-rank approximations of $n$-IFC could be game changing. There is growing interest in such dimensionality reduction approaches, including tensor network states for many-body wave functions~\cite{rmp-tns,tns-2}, tensor train for differential equations~\cite{tt-decomp-2011,TT-feynman-2022,TT-for-fluid-dynamics-2024} and Feynman diagrams~\cite{TT-feynman-2022,multiscale-space-time-prx}, tensor hyper-contraction~\cite{thc-1,thc-2,thc-3} and density fitting~\cite{df-qchem} for electron interactions in quantum chemistry, and feature optimization for atomic machine learning~\cite{Feature-optimization-2018,feature-selection-2018,feature-selections-atomic-ml-2021,Darby2022}. 
\mbox{Recent} work has employed singular value decomposition to compress electron-phonon ($e$-ph) interactions and greatly speed up first-principles $e$-ph calculations~\cite{compress-eph}. 
For ph-ph interactions, previous work has taken advantage of crystal symmetry to reduce the number of free parameters and compute the $n$-IFCs with fewer DFT force calculations~\cite{Group-theoretical-IFC-Marianetti,symmetry-ifc}. 
The compressed sensing technique has also been used to obtain $n$-IFCs using sparse solvers~\cite{compress-sensing-for-Anharmonicity,compress-sensing-for-Anharmonicity-prb}.
However, in previous work the $n$-IFCs are still handled explicitly in full tensor form, $\Phi_{ijk}$ and $\Phi_{ijkl}$ for 3- and 4-IFCs respectively, and a low-rank representation of $n$-IFCs is still missing. 
\\
\indent
In this Letter, we introduce a low-rank tensor ansatz for $n$-ph interactions in momentum space based on the CANDECOMP/PARAFAC (CP) decomposition~\cite{cp-decomp}, a compression method used in tensor learning~\cite{CP-learning-1,CP-learning-2,CPD-MP2-2025-arxiv}. 
The proposed ansatz is a permanent CP (PCP) decomposition, which generalizes the CP decomposition 
to enforce bosonic statistics. 
To find an optimal low-rank PCP decomposition, we formulate the tensor learning problem and solve it on GPU hardware.   
\mbox{The optimized} low-rank PCP tensors achieve large compression factors of 10$^3$--10$^4$ with minimal compression losses of only a few percent in all materials we study. 
This result reveals the inherent low-dimensionality of $n$-ph interactions and enables a speedup of nearly three orders of magnitude for calculations using $n$-IFCs, including phonon relaxation times and thermal conductivity. 
We also introduce constraints to treat ph-ph interactions for long-wavelength acoustic phonons, leading to nearly lossless predictions of thermal conductivity compared to calculations using full $n$-IFC tensors. 
Finally, we show that the PCP ansatz can uncover dominant modes in $n$-ph interactions, providing a valuable tool for understanding microscopic thermal transport mechanisms and formulating accurate minimal models. 
Beyond ph-ph interactions, the PCP ansatz and corresponding open-source routines developed here provide a blueprint for modeling momentum-dependent tensors, with broad applications in condensed matter physics. 
\\
\indent
We consider the $n$-ph interaction $V^{(n)}(\QQ_1,...,\QQ_n)$, which describes the scattering amplitude of $n$ phonon modes. Each mode is specified by a wave vector 
$\mathbf{q}_i$ and branch index $\nu_i$, collectively denoted as 
$\mathbf{Q}_i = (\nu_i, \mathbf{q}_i)$ for $i = 1,\ldots,n$. For instance, the 3-ph interaction is given by   
\begin{align}\label{eq:V3}
        V^{(3)}(&\QQ_1,\QQ_2,\QQ_3) =  
        \delta\left(\sum_{i=1}^{3}\qq_i\right)
        \sum_{b_1,l_2b_2,l_3b_3} \sum_{\alpha_1\alpha_2\alpha_3} \nonumber \\
        &
        \times\Phi^{\alpha_1,\alpha_2,\alpha_3}_{0b_1,l_2b_2,l_3b_3}\frac{e^{\QQ_1}_{\alpha_1 b_1}e^{\QQ_2}_{\alpha_2 b_2}e^{\QQ_3}_{\alpha_3 b_3}}{\sqrt{m_{b_1}m_{b_2}m_{b_3}}} e^{i\qq_2 \bm{r}_{l_2} + i\qq_3\bm{r}_{l_3}},
\end{align}
where each of the $l$, $b$, and $\alpha$ label the primitive cell, atom, and Cartesian coordinate, respectively; $e^{\QQ}_{\alpha b}$ is the  displacement eigenvector of phonon mode $\QQ$, $m_b$ is the mass of atom $b$, and $\bm{r}_{l}$ is the position of primitive \mbox{cell $l$.}
Above, $\Phi^{\alpha,\alpha_1,\alpha_2}_{0b,l_1b_1,l_2b_2}$ is the \mbox{3-IFC} tensor associated with three atomic displacements, the first displacement fixed at the cell origin.
For general $n$-ph interactions, analogous $n$-IFC tensors $\bm{\Phi}^{(n)}$ can be defined. 
\\
\indent
To compress the $n$-ph interactions $V^{(n)}$, we propose the PCP decomposition as the low-rank ansatz
\begin{align}\label{eq1:V}
    \tilde{V}^{(n)}&\left(\QQ_1,...,\QQ_n\right) = \delta\left(\sum_{i=1}^{n}\qq_i\right)  \nonumber 
    \\
    &\times \sum_{\xi=1}^{N^{(n)}_c} 
    \frac{\lambda_\xi}{n!} 
    \sum_{\sigma \in S_n}  A^\xi_{1}(\QQ_{\sigma_1})A^\xi_{2}(\QQ_{\sigma_2})...A^\xi_{n}(\QQ_{\sigma_n}), 
\end{align}
where $N^{(n)}_c$ is the PCP rank, $\lambda_\xi$ the $\xi$-th PCP singular value, $A_i^\xi$ are the corresponding PCP modes, and $S_n$ is the symmetric group of order $n$ containing all the permutations of $\left(1,...,n\right)$.
Note that Eq.~(\ref{eq1:V}) correctly preserves the bosonic exchange symmetry and momentum conservation of $n$-ph interactions. 
\\
\indent
\textcolor{black}{
The $n$-ph interactions in compressed form involve $n$ momenta $(\QQ_1,\QQ_2, ..., \QQ_n)$, each associated with $N_c^{\left(n\right)}$ PCP modes $A_\xi\left(\QQ\right)$, which are functions defined on a grid of size $N_{\QQ}$. This gives a storage requirement of $\mathcal{O}(nN_c^{\left(n\right)}N_{\QQ})$ for the $n$-ph interactions in compressed form.
For example, storing 3-ph interactions $V^{(3)}$ conventionally} requires $\mathcal{O}(N_{\QQ}^{2})$ resources versus $\mathcal{O}(3 N_c^{(3)}N_{\QQ})$ for its PCP compressed counterpart $\tilde{V}^{(3)}$. 
This yields a large storage reduction, by a factor of $N_{\QQ}/(3N_c^{(3)})\approx$ 10$^3$--10$^4$ in a typical calculation.
%
%

Since $V^{(n)}$ is obtained by transforming the $n$-IFC $\bm{\Phi}^{(n)}$ to momentum space, we derive the corresponding low-rank ansatz for 3- and 4-IFCs in real space: 
\begin{align}\label{eq2:phi3}
&\tilde{\Phi}^{\alpha_1,\alpha_2,\alpha_3}_{l_1b_1,l_2b_2,l_3b_3} = \sum_{\xi=1}^{N^{(3)}_c} 
    \frac{\lambda_\xi}{3!}\sum_{\sigma \in S_3} 
   \sum_{l}\prod_{i=1}^3 A^\xi_{\sigma_i}(l_i-l,b_i\alpha_i),
   \\
   &\tilde{\Phi}^{\alpha_1,\alpha_2,\alpha_3,\alpha_4}_{l_1b_1,l_2b_2,l_3b_3,l_4b_4} = \sum_{\xi=1}^{N^{(4)}_c} 
    \frac{\lambda_\xi}{4!}\sum_{\sigma \in S_4} 
   \sum_{l}\prod_{i=1}^4 B^\xi_{\sigma_i}(l_i-l,b_i\alpha_i),
   \nonumber
\end{align}
where $A^\xi_{i}(l,b\alpha)$ and $B^\xi_{i}(l,b\alpha)$ are real-space representations of $A^\xi_{i}(\QQ)$ in Eq.~\eqref{eq1:V} for 3-ph and 4-ph interactions respectively.
To enforce the acoustic sum rule (ASR)~\cite{symmetry-ifc} \textcolor{black}{on the compressed $n$-ph interactions  $\tilde{V}^{(n)}$, we impose $\sum_{l,b} A^\xi_{i}(l,b\alpha) = 0$, which preserves the ASR in the PCP compression.}
Derivations of $n$-ph interactions and their PCP decomposition are provided in the Supplemental Material (SM)~\cite{sm}. 
The computational cost of PCP-compressed $\tilde{\bm{\Phi}}^{(n)}$ and $\tilde{V}^{(n)}$ scales linearly with system size, making it promising for calculations in complex materials with large unit cells.

To find optimal PCP modes $A^\xi_i$ (or $B^\xi_i$) that best approximate the original IFC tensors $\bm{\Phi}^{(n)}$ computed with DFT, we minimize the loss function: 
\begin{equation}\label{eq:loss}
    L = \|\tilde{\bm{\Phi}}^{(n)}\left[ \bm{A} \right]-\bm{\Phi}^{(n)}\|^2,
\end{equation}
where $\tilde{\bm{\Phi}}^{(n)}$ is the low-rank IFC tensor in Eq.~(\ref{eq2:phi3}), and the norm is defined as \mbox{$\| \bm{x} \| = \sqrt{\sum_{I} x_I^2}$}~
\footnote{In the optimization, $A^\xi_i(l,b\alpha) $ is truncated in real space up to a cutoff $r_c$ large enough for the low-rank IFC tensor $\tilde{\bm{\Phi}}^{(n)}$ to span all nonzero elements in the full IFC tensor $\bm{\Phi}^{(n)}$.}.
\begin{figure}
    \centering
    \includegraphics[width=1.0\linewidth]{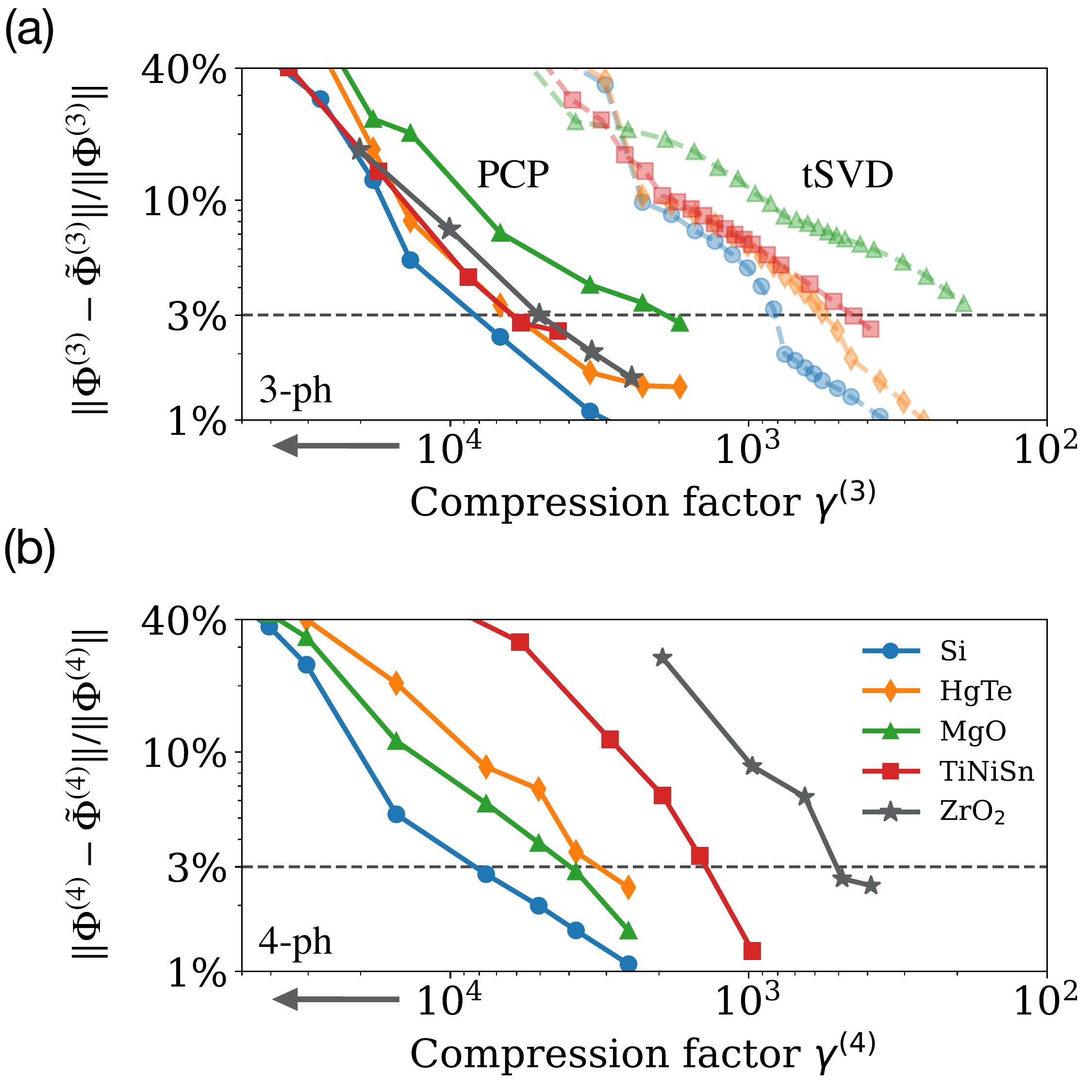}
    \caption{Relative compression loss vs. compression factor $\gamma^{(n)}$. 
    (a) Results for 3-IFC tensors comparing PCP (solid lines) and tSVD (dashed lines). 
    (b) Results for 4-IFC tensors.
    The compression factor $\gamma^{(n)}$ is defined in Eq.~(\ref{eq:CR}).
   }
    \label{Fig1:error}
\end{figure}
%
%
This is a typical tensor learning problem that lacks closed-form solutions for high-order tensors.
In addition, existing optimization algorithms are computationally demanding for large tensors.
We address this challenge by developing a computational toolkit, called {\sc Phonon-PCP}, implemented using \mbox{{\sc PyTorch}}~\cite{pytorch}. 
This optimization routine follows a standard ``neural network"-like training loop, where $\tilde{\bm{\Phi}}^{(n)}$ is evaluated in a forward pass, and $A^\xi_i$ is the learnable weight trained using automatic differentiation.  
A typical optimization involves ${10^3}-{10^4}$ trainable parameters, depending on the material. 
\textcolor{black}{
The algorithm runs efficiently on GPUs – the compression process converges in a few minutes (Si and other simple systems) to one hour (ZrO$_2$) on a single NVIDIA A100 chip.
}
\\
\indent
We apply our approach to five materials: Si, HgTe, MgO, TiNiSn, \textcolor{black}{and monoclinic ZrO$_2$}. 
\textcolor{black}{These materials span a wide range of properties, including nonpolar and polar (or ionic) bonds, high and low crystal symmetry, and different levels of anharmonicity.}
For each material, we generate force-displacement datasets from DFT force calculations on supercells and then employ {\sc Alamode}~\cite{alamode} to extract the 3- and 4-IFCs 
\footnote{We include 3-IFCs up to the fifth-nearest neighbors and 4-IFCs up to the second-nearest neighbors for Si, HgTe and MgO. 
For \mbox{TiNiSn}, the 3-IFCs are truncated to 6.4 \AA \ and the 4-IFCs to 3.8 \AA. For \mbox{ZrO$_2$}, the 3-IFCs are truncated to 5.3 \AA \ and the 4-IFCs to 2.7 \AA.}.
\begin{figure}
    \centering
    \includegraphics[width=1.0\linewidth]{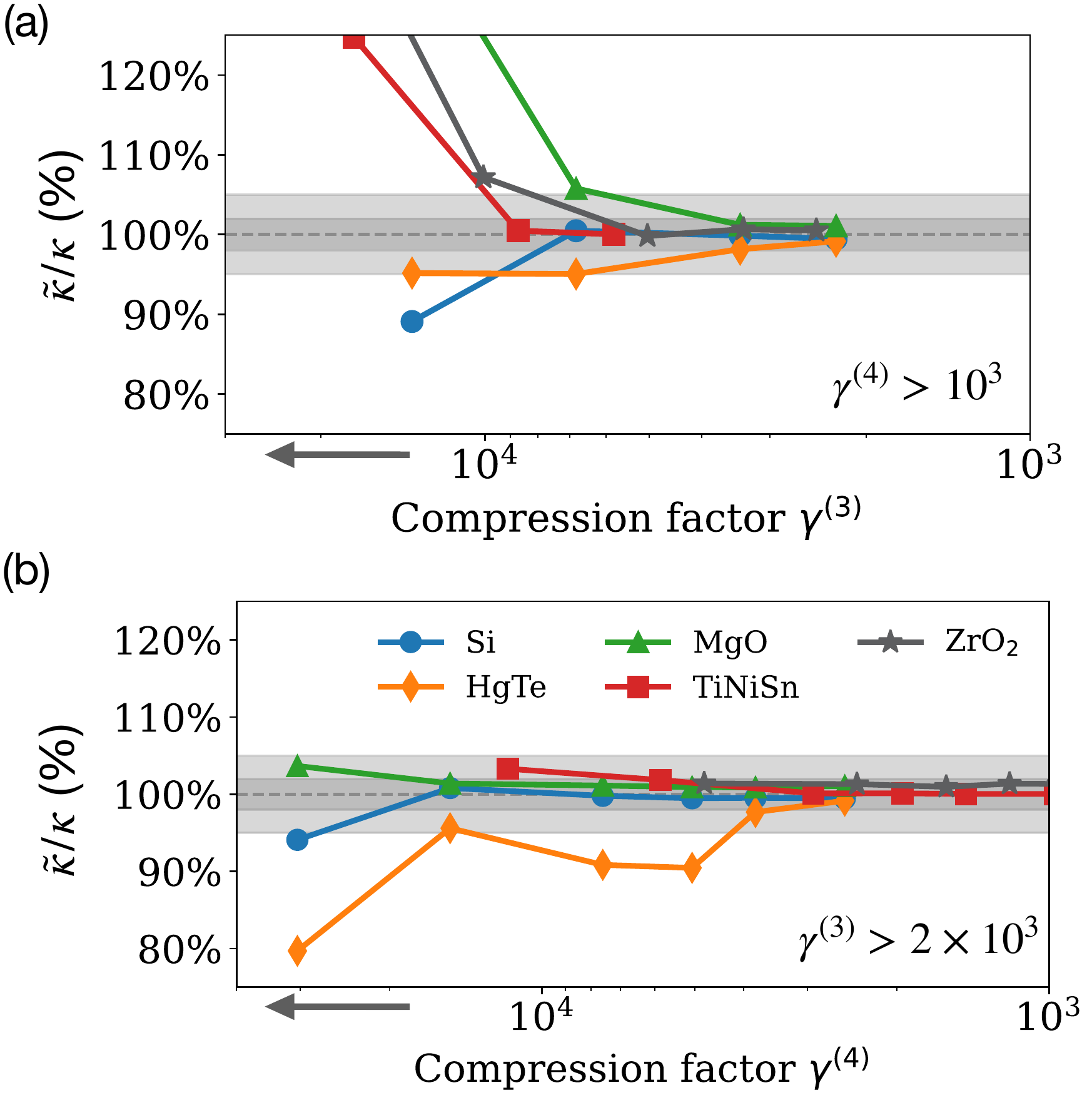}
    \caption{Ratio of the thermal conductivity $\tilde{\kappa}$, computed using compressed 3- and 4-ph interactions, to the thermal conductivity $\kappa$ computed with uncompressed tensors \textcolor{black}{at \mbox{300 K}}. Results are shown as a function of: (a) 3-ph compression factor, $\gamma^{(3)}$, for fixed $\gamma^{(4)}\!>\!10^3$ ($N_c^{(4)} \!=\! 48$), and (b) 4-ph compression factor, $\gamma^{(4)}$, for fixed $\gamma^{(3)}\!>\!2\times10^3$ ($N_c^{(3)} \!=\! 24$ \textcolor{black}{for all materials except ZrO$_2$, and $N_c^{(3)} \!=\! 144$ for ZrO$_2$}).
    The light (dark) shaded regions show the $95\%$ ($98\%$) accuracy windows.
    }
    \label{fig:thermal-relative-error}
\end{figure}
\textcolor{black}{The symmetry of the $n$-IFCs is taken into account in the force constant fitting process~\cite{alamode}.}
We calculate DFT forces using VASP~\cite{vasp,vasp-2} with PBEsol functional~\cite{pbesol}. 
For polar materials, we calculate Born effective charges using density functional perturbation theory in VASP, and use the Ewald summation to compute the nonanalytic part of the dynamical matrix~\cite{Ewald}.
Additional computational details, 
are provided in the SM~\cite{sm}. 
\begin{figure}
    \centering
    \includegraphics[width=1.0\linewidth]{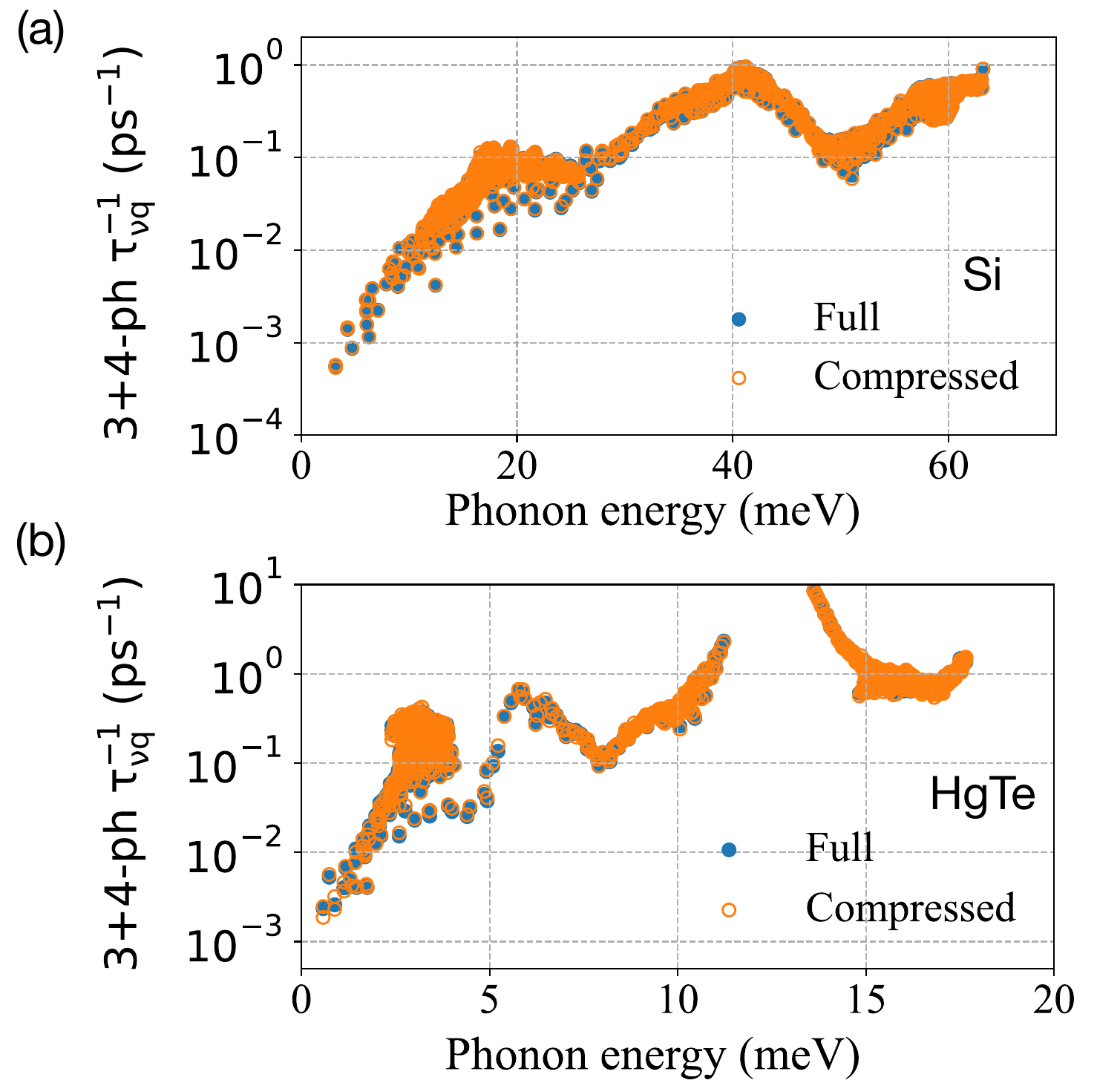}
    \caption{Phonon scattering rates combining 3- and 4-ph interactions, shown as a function of phonon energy for (a) Si and (b) HgTe, comparing results for full IFC tensors (blue dots) and compressed IFC tensors (orange circles). In both materials, we show results for compression factors $\gamma^{(3)} = 2200$ for 3-ph, and $\gamma^{(4)} = 2500$ for 4-ph interactions.
    }
    \label{Fig2:theraml-cond}
\end{figure}
\\
\indent
To assess the accuracy of PCP for compressing \mbox{$n$-ph} interactions, we calculate the error relative to uncompressed $n$-IFCs, expressed as the compression loss ${\|\tilde{\bm{\Phi}}^{(n)}-\bm{\Phi}^{(n)}\|}/{\|\bm{\Phi}^{(n)}\|}$, as a function of the compression factor $\gamma^{(n)}$, which quantifies parameter reduction in the $n$-IFCs: 
\begin{equation}\label{eq:CR}
    \gamma^{(n)} = \frac{\text{\# of nonzero entries of} \ \bm{\Phi}^{(n)}}{N_c^{(n)}}.
\end{equation}
This error analysis is shown for the 3-ph interactions in Fig.~\ref{Fig1:error}(a). 
We achieve compression factors of $10^3$--$10^4$ at the 3\% compression error threshold. 
For comparison, a truncated singular value decomposition (tSVD) of 3-ph interactions, inspired by our previous work on $e$-ph coupling~\cite{compress-eph}, can only reach compression factors of a few hundred for the same 3\% compression loss. \mbox{Therefore,} the
PCP approach provides a 10-times more favorable dimensionality reduction than SVD for all materials we study. 
We attribute the superior performance of PCP to its flexible ansatz combined with the correct permutation symmetry and the explicit parameter optimization during tensor learning, which enable effective low-rank representation of $n$-ph interactions.
We achieve similarly large compression factors for 4-ph interactions, as shown in Fig.~\ref{Fig1:error}(b). These results highlight the inherent low-dimensionality of $n$-ph interactions and show their accurate low-rank approximation using tensor learning. 
\\
\indent
\textcolor{black}{Note that crystal symmetry can also reduce the size of the $n$-IFCs, but it cannot be used to speed up calculations of phonon relaxation times and thermal conductivity, which require the full $n$-IFCs. In addition, the size reduction from symmetry is  significantly smaller than the PCP compression factor shown here~\cite{sm}.}
\\
\indent
Using compressed $n$-ph interactions, we compute the thermal conductivity at 300 K for the four materials. Our calculations use the single-mode relaxation time approximation~\cite{single-mode-approximation,peierls1955,ziman1960} and include both 3- and 4-ph interactions~\cite{ph-scattering-pr-1962,ph-scattering-pr-1974,ph-scattering-prb-1983,ph-scattering-prb-2016} with a converged number of scattering processes for 3- and 4-ph scattering rates~\cite{sampling-technique-phscat} (see details in SM~\cite{sm}). 
\begin{figure}
    \centering
    \includegraphics[width=0.9\linewidth]{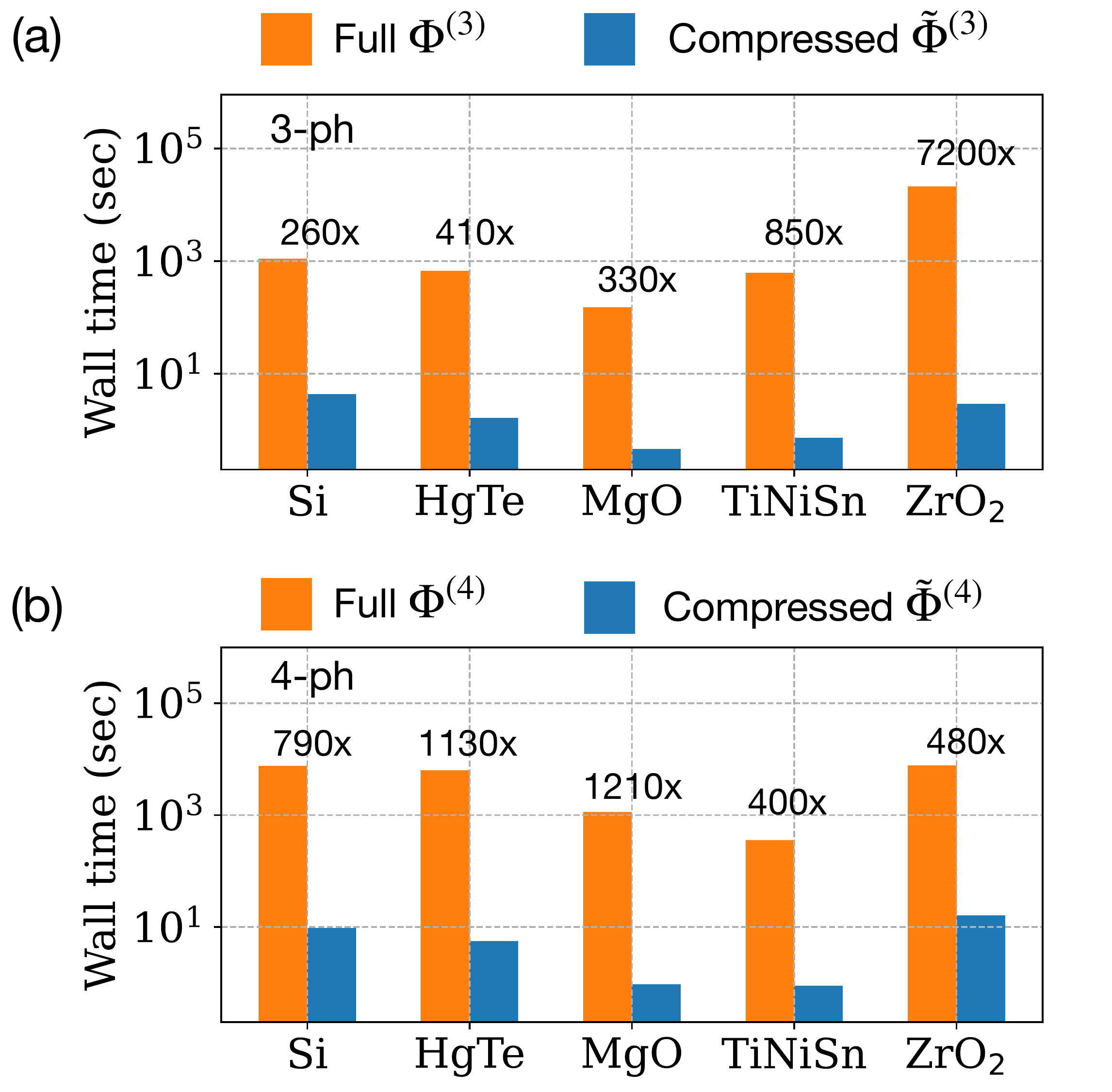}
    \caption{Comparison of computational cost (CPU wall time) for calculations \textcolor{black}{of phonon scattering rates} using compressed and full IFC tensors. Results are shown for (a) 3-ph scattering rates with compression factors $\gamma^{(3)} \!>\! 2200$ and (b) 4-ph scattering rates with compression factors $\gamma^{(4)} \!>\! 1000$ in all materials. 
    The speedup achieved with compressed tensors is given in figure for each material.\vspace{-10pt}
    }
    \label{Fig3:timing}
\end{figure}
The thermal conductivity $\tilde{\kappa}$, computed with compressed $n$-IFC tensors, is compared with the reference value $\kappa$ from full (uncompressed) tensors in Fig.~\ref{fig:thermal-relative-error}(a) and (b). 
We find that $\tilde{\kappa}$ converges to the reference value for large compression factors, $\gamma^{(3)}\!\approx\!7\times10^{3}$ and $\gamma^{(4)}\!\approx\!10^{4}$ for Si, MgO, TiNiSn, \textcolor{black}{and ZrO$_2$}, and slightly smaller $\gamma^{(4)} \!=\! 3\times10^3$ for HgTe, where the 4-ph contribution is more important~\cite{thermal-trans-2}. 
%
Even for these very large compression factors, the approximate thermal conductivity is still within 2\% of the reference value obtained with full IFCs~\cite{sm}.
This high accuracy for thermal conductivity is a result of the low compression losses for $n$-IFCs. 
\\
\indent 
We also examine the accuracy of compressed IFC tensors for calculating the microscopic scattering rate for each phonon mode, $\tau_{\nu\qq}^{-1}$. 
In Figs.~\ref{Fig2:theraml-cond}(a) and (b), the phonon scattering rates from compressed IFC tensors are nearly identical to those from full IFC tensors for each single phonon mode, even in cases where $\tau_{\nu\qq}^{-1}$ varies by up to four orders of magnitude over the phonon spectrum. 
We calculate the coefficient of determination $R^{2}$~\cite{regression-book} between approximate and reference $\tau_{\nu\qq}^{-1}$, 
and find values of $R^{2}=0.9996$ for Si and $R^{2}=0.9986$ for HgTe, further confirming the accuracy of the compressed $n$-ph interactions. 
The approximate scattering rates achieve a similar accuracy in TiNiSn and MgO~\cite{sm}. 
\\
\indent
The dimensionality reduction provided by PCP leads to massive cost savings for calculations involving $n$-ph interactions, with speed-up proportional to the compression factor. 
In Fig.~\ref{Fig3:timing}, we compare CPU wall times for calculations of 3-ph and 4-ph scattering rates using full and compressed $n$-IFCs with large compression factors ($\gamma^{(3)} \!>\! 2200$ and $\gamma^{(4)} \!>\! 1000$).  
For all materials studied here, the use of compressed IFC tensors enables a speed-up of 260--\textcolor{black}{7200} in calculations of thermal conductivity. 
Calculations of $\tau_{\nu\qq}^{-1}$ using compressed IFC tensors are so efficient that 
\textcolor{black}{computing the thermal conductivity with 3- and 4-ph interactions takes less than a minute for each material studied here, including monoclinic ZrO$_2$ with a 12-atom unit cell.} 
Using compressed IFC tensors, we are able to accurately extrapolate the thermal conductivity to the thermodynamic limit (TDL) by employing progressively denser momentum grids. The importance of extrapolating to the TDL is clear from the case of HgTe, where the TDL-extrapolated $\kappa$ is 47\% larger than the value obtained with reasonable grid sizes of $16^3$~\cite{sm}. 
The calculation for the largest grid, $250^3$, requires only 2.5 CPU node-hours. 
\textcolor{black}{We attribute the slow convergence of the thermal conductivity in HgTe to the important role of 4-ph processes (see analysis in SM~\cite{sm}).}
\\
\indent
\textcolor{black}{In Fig. \ref{Fig3:timing}(a), the speed-up increases for increasing unit cell sizes. For large unit cells, such as in ZrO$_2$, the speed-up becomes comparable to the compression factor because the cost of computing the $n$-ph interactions dominates the total CPU wall time~\cite{sm}.}
\textcolor{black}{The significant speed-up achieved by PCP can be understood by comparing the $n$-ph interactions and their compressed counterparts.
The size of the $n$-IFC tensor in real space, $\Phi^{(n)}$, scales as $N_R^{n-1}(3N_a)^n$, where $N_R$ is the number of Wigner-Seitz cells, which is the same as the size of the coarse $\mathbf{q}$-point grid, and $N_a$ is the number of atoms in the unit cell, so that 3$N_a$ is the number of phonon modes. 
The summation over these variables in the conventional $n$-ph interactions, exemplified by the 3-ph case in Eq.~\eqref{eq:V3}, is replaced by a summation over $N_c^{\left(n\right)}$ PCP modes for the compressed $n$-ph interactions in Eq.~\eqref{eq1:V}. 
This reduces the computational cost of $V^{\left(n\right)}(\QQ_1,\ \ldots,\ \QQ_n)$  by a factor of $\mathcal{O}\left({3N}_a^nN_R^{n-1}/N^{(n)}_c\right)$ (see analysis in SM~\cite{sm}).}
\\
\indent 
The PCP decomposition also offers interesting ways to analyze $n$-ph interactions. 
We express the $n$-ph anharmonic energy $E^{(n)}(\bm{u})$ in terms of PCP modes~\cite{sm}, 
\begin{align}
    E^{(n)}(\bm{u}) &= \sum_{l}\sum_{\xi}\frac{\lambda_\xi}{n!} \prod_{i=1}^n 
    \phi^\xi_i(l,\bm{u}) , \label{eq:En}
    \\
    \phi^\xi_i(l,\bm{u}) &= \braket{\bm{A}^\xi_i(l), \bm{u}}  \label{eq:pcp_mode}
    \\ 
    &= \sum_{l'b\alpha} A^\xi_{\sigma_i}(l'-l,b\alpha) u(l'b\alpha),
    \nonumber
\end{align}
where $u(lb\alpha)$ is the displacement of atom $b$ in primitive cell $l$ along the $\alpha$ direction. 
Here, the PCP modes \mbox{$[\bm{A}^\xi_i(l)]_{l'b\alpha} = A^\xi_{\sigma_i}(l'-l,b\alpha)$} are viewed as local vibrational modes, centered in the $l$-th primitive cell, which give dominant contributions the $n$-th order anharmonic energy $E^{(n)}(\bm{u})$.
Therefore, the projection $\phi^\xi_i(l,\bm{u})$ defined in Eq.~\eqref{eq:pcp_mode} is a descriptor of the atomic environment that quantifies the similarity between the local atomic displacement $\bm{u}$ and the PCP modes $\bm{A}^\xi_i(l)$. (Interestingly, the anharmonic energy for PCP decomposition in Eq.~(\ref{eq:En}) has a structure similar to the slave mode~\cite{slave-mode-expansion,slave-mode-2} and the atomic cluster expansion models~\cite{ACE-prb-2019}.)

\begin{figure}
    \centering
    \includegraphics[width=1.0\linewidth]{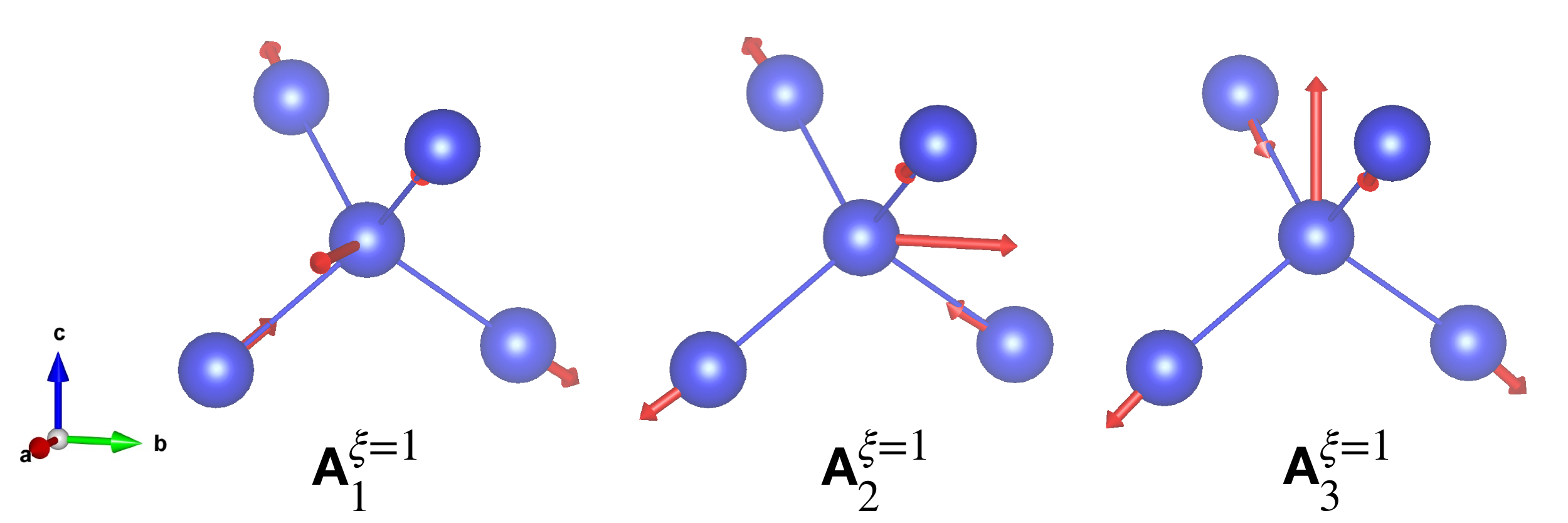}
    \caption{The PCP mode triplet $A^{\xi=1}_{i=1,2,3}$ with the largest 3-ph coupling strength in Si. The three modes are related by the C$_3$ rotation symmetry along the [111] direction in Si. }
    \label{Fig4:modes}
\end{figure}

\textcolor{black}{These PCP modes can be viewed as generalized eigenvectors of the $n$-ph interaction tensor, and provide direct physical information about the dominant vibrational patterns.} 
In Fig.~\ref{Fig4:modes}, we visualize the three modes associated with the largest PCP singular value for 3-ph interactions in Si.
The PCP-mode triplet \{$A^{\xi=1}_{1}, A^{\xi=1}_{2}, A^{\xi=1}_{3}$\} provides the best rank-1 approximation of 3-IFC tensors in Si and is obtained by setting the PCP rank to $N_c^{(3)}$ \!=\! 1 in the training process. 
These three modes are related by a $C_3$ rotation along the [111] crystal direction: 
\begin{equation}
    \hat{C}_{3}A^{\xi=1}_{1} = A^{\xi=1}_{2},\  \hat{C}_{3}A^{\xi=1}_{2} = A^{\xi=1}_{3},\ \hat{C}_{3}A^{\xi=1}_{3} = A^{\xi=1}_{1}. 
    \nonumber
\end{equation}
Even though we do not explicitly preserve the space-group symmetry when generating the PCP modes, the rank-1 PCP ansatz is capable of learning the $C_3$ symmetry of Si encoded in the uncompressed 3-IFC tensors during the optimization process. 
In general, the crystal symmetry is preserved in the compressed IFCs, with only a small symmetry loss resulting from compression (see SM~\cite{sm}). 

In summary, we propose a tensor decomposition of \mbox{$n$-ph} interactions and show compression of $n$-IFC tensors using GPU-accelerated tensor learning. 
Our PCP decomposition reveals the inherent low-dimensionality of 3- and 4-ph interactions in crystals \textcolor{black}{with generic symmetry and unit cell size}, enabling compression factors greater than $10^3$ for minimal compression errors of less than 3\%. 
We achieve corresponding cost savings for calculations of phonon scattering rates and thermal conductivity. 
The large speed-up makes our method suitable for high-throughput screening of thermal transport in materials and offers a promising pathway to go beyond 3- and 4-ph interactions. 
\textcolor{black}{The PCP-compressed $n$-ph interactions can be combined with CPU and GPU parallelization to accelerate modeling of ph-ph interactions.}
The PCP decomposition can also uncover dominant atomic environment descriptors, providing valuable information for formulating machine learning atomic force fields. Future work will explore \mbox{5-ph} and higher $n$-ph interactions in strongly anharmonic materials, use PCP to obtain simplified coupled-mode equations for nonlinear phonon processes, and accelerate simulations of ultrafast phonon dynamics.

\vspace{10pt}
\begin{acknowledgments}
Y.L. thanks Junjie Yang for fruitful discussions. 
This work was supported by the National Science Foundation under Grant No. OAC-2209262. 
Y.L. acknowledges support from the Eddleman Fellowship.
This research used resources of the National Energy Research Scientific Computing Center, a DOE Office of Science User Facility supported by the Office of Science of the U.S. Department of Energy under Contract No. DE-AC02-05CH11231 using NERSC award NERSC DDR-ERCAP0026831.
\end{acknowledgments}

\end{document}


\title{Supplemental Material for \\``Tensor Learning and Compression of N-phonon Interactions''}
\author{Yao Luo}%
\affiliation{Department of Applied Physics and Materials Science, and Department of Physics, California Institute of Technology, Pasadena, California 91125, USA}

\author{Dhruv Mangtani}%
\affiliation{Department of Applied Physics and Materials Science, and Department of Physics, California Institute of Technology, Pasadena, California 91125, USA}

\author{Shiyu Peng}%
\affiliation{Department of Applied Physics and Materials Science, and Department of Physics, California Institute of Technology, Pasadena, California 91125, USA}

\author{Jia Yao}%
\affiliation{Department of Applied Physics and Materials Science, and Department of Physics, California Institute of Technology, Pasadena, California 91125, USA}

\author{Sergei Kliavinek}%
\affiliation{Department of Applied Physics and Materials Science, and Department of Physics, California Institute of Technology, Pasadena, California 91125, USA}

\author{Marco Bernardi}
\affiliation{Department of Applied Physics and Materials Science, and Department of Physics, California Institute of Technology, Pasadena, California 91125, USA}
\email{bmarco@caltech.edu}

\maketitle
\newpage
\vspace{-10pt}

\section{Derivation of $n$-ph interactions and their PCP decomposition}
The $n$-IFC tensor in PCP format is
\begin{align}
\label{eq:phi}
\tilde{\Phi}^{\alpha_1,...,\alpha_n}_{l_1b_1,...,l_nb_n} = \sum_{\xi=1}^{N^{(n)}_c} 
    \frac{\lambda_\xi}{n!}\sum_{\sigma \in S_n} 
   \sum_{l}\prod_{i=1}^n A^\xi_{\sigma_i}(l_i-l,b_i\alpha_i),
\end{align}
where $l$, $b$, and $\alpha$ index the primitive cell, basis atom, and Cartesian coordinate, respectively.
In the following, we show that this expression is consistent with the PCP decomposition of $n$-ph interactions, $\tilde{V}^{(n)}$, in Eq.~(2) of main text.

The compressed $n$-ph interaction in momentum space, $\tilde{V}^{(n)}$, is obtained from a basis transformation of $\tilde{\Phi}$: 
\begin{align}
\label{eq:V-ft-phi}
       \tilde{V}^{(n)}(\QQ_1,...,\QQ_n) =  
       \frac{1}{N} \sum_{l_1b_1,\alpha_1} \frac{e^{\QQ_1}_{\alpha_1 b_1}}{\sqrt{m_{b_1}}}e^{i\qq_1 \bm{r}_{l_1}} 
        ...
         \sum_{l_nb_n,\alpha_n} \frac{e^{\QQ_n}_{\alpha_n b_n}}{\sqrt{m_{b_n}}}e^{i\qq_n \bm{r}_{l_n}} 
\tilde{\Phi}^{\alpha_1,...,\alpha_n}_{l_1b_1,...,l_nb_n} ,
\end{align}
where $N$ is the number of primitive cells in a Born–von Kármán (BvK) supercell. 
Substituting Eq.~\eqref{eq:phi} into Eq.~\eqref{eq:V-ft-phi}, and exchanging the order of $\prod$ and $\sum$, we get  
\begin{align}
\label{eq:trick}
    \tilde{V}^{(n)}(\QQ_1,...,\QQ_n) &=  
    \sum_{\xi=1}^{N^{(n)}_c}\frac{\lambda_\xi}{n!}\sum_{\sigma \in S_n} \frac{1}{N} \sum_{l}\prod_{i=1}^{n}  \left (\sum_{l_ib_i,\alpha_i} \frac{e^{\QQ_1}_{\alpha_i b_i}}{\sqrt{m_{b_i}}}e^{i\qq_i \bm{r}_{l_i}} A_{\sigma_i}(l_i-l,b_i\alpha_i) \right).
\end{align}
Let us define the PCP modes in momentum space, 
$$  A^{\xi}_{\sigma_i}(\QQ) = \sum_{l_i'b_i,\alpha_i} \frac{e^{\QQ_1}_{\alpha_i b_i}}{\sqrt{m_{b_1}}}e^{i\qq_i\bm{r}_{l_i'}} A_{\sigma_i}(l_i',b_i\alpha_i).
$$
After the change of variable $l_i'=l_i-l$, we obtain 
\begin{align}
\tilde{V}^{(n)}(\QQ_1,...,\QQ_n) &=  
    \sum_{\xi=1}^{N^{(n)}_c}\frac{\lambda_\xi}{n!}\sum_{\sigma \in S_n} \frac{1}{N} \sum_{l}\prod_{i=1}^{n}  \left (e^{i\qq_i\bm{r}_{l}}A^{\xi}_{\sigma_i}(\QQ_i)  \right)
    \\
    &=    \sum_{\xi=1}^{N^{(n)}_c}\frac{\lambda_\xi}{n!}\sum_{\sigma \in S_n} \frac{1}{N} \sum_{l} e^{i\sum_{i=1}^n\qq_i\bm{r}_{l}} \prod_{i=1}^{n}  A^{\xi}_{\sigma_i}(\QQ_i)
        \\
    &=    \sum_{\xi=1}^{N^{(n)}_c}\frac{\lambda_\xi}{n!}\sum_{\sigma \in S_n} \delta\left(\sum_{i=1}^n\qq_i\right) \prod_{i=1}^{n} A^{\xi}_{\sigma_i}(\QQ_i).
\end{align}
Using the property that a permanent is invariant under matrix transpose, we exchange $i$ and $\sigma_i$ in $A^{\xi}_{\sigma_i}(\QQ_i)$, and obtain the $n$-ph interactions in momentum space in PCP format:
\begin{align}
    \tilde{V}^{(n)}(\QQ_1,...,\QQ_n) =  \delta\left(\sum_{i=1}^n\qq_i\right) \sum_{\xi=1}^{N^{(n)}_c}\frac{\lambda_\xi}{n!}\sum_{\sigma \in S_n} \prod_{i=1}^{n} A^{\xi}_{i}(\QQ_{\sigma_i}),
\end{align}
which is Eq. (2) in the main text.

\section{Anharmonic energy $E^{(n)}$ in PCP decomposition}
Using the definition of the $n$-IFCs, the $n$th-order anharmonic energy $E^{(n)}(\bm{u})$ is:
\begin{equation}
    E^{(n)}(\bm{u}) = \frac{1}{n!}\sum_{l_1 b_1 \alpha_1}...\sum_{l_n b_n \alpha_n} \tilde{\Phi}^{\alpha_1,...,\alpha_n}_{l_1b_1,...,l_nb_n} 
    u(l_1b_1\alpha_1)u(l_2b_2\alpha_2)...u(l_nb_n\alpha_n). 
\end{equation}
Substituting Eq.~\eqref{eq:phi} into the equation above, we obtain
\begin{equation}
   E^{(n)}(\bm{u}) = \frac{1}{n!}\sum_{l_1 b_1 \alpha_1}u(l_1b_1\alpha_1)...\sum_{l_n b_n \alpha_n} u(l_nb_n\alpha_n) \sum_{\xi=1}^{N^{(n)}_c} 
    \frac{\lambda_\xi}{n!}\sum_{\sigma \in S_n} 
   \sum_{l}\prod_{i=1}^n A^\xi_{\sigma_i}(l_i-l,b_i\alpha_i).
\end{equation}
Exchanging the order of $\prod$ and $\sum$ in Eq.~(\ref{eq:trick}), we rewrite this equation as
\begin{equation}
       E^{(n)}(\bm{u}) = \sum_{l}\sum_{\xi=1}^{N^{(n)}_c} 
    \frac{\lambda_\xi}{n!} \frac{1}{n!}  \sum_{\sigma \in S_n} 
   \prod_{i=1}^n \left(\sum_{l_i b_i \alpha_i} A^\xi_{\sigma_i}(l_i-l,b_i\alpha_i)u(l_ib_i\alpha_i)\right).
\end{equation}
To simplify this expression, we define the projections appearing in Eq.~(7) of main text:
\begin{align}
    \phi^\xi_i(l,\bm{u}) &= \braket{\bm{A}^\xi_i(l), \bm{u}}  \label{eq:pcp_mode}
    = \sum_{l'b\alpha} A^\xi_{\sigma_i}(l'-l,b\alpha) u(l'b\alpha).
\end{align}
In terms of these quantities, the $n$th-order anharmonic energy $E^{(n)}(\bm{u})$ becomes
\begin{equation}
    E^{(n)}(\bm{u}) = \sum_{l}\sum_{\xi=1}^{N^{(n)}_c} 
    \frac{\lambda_\xi}{n!} \frac{1}{n!}  \sum_{\sigma \in S_n} 
   \prod_{i=1}^n \phi^\xi_{\sigma_i}(l,\bm{u}).
\end{equation}
Since $\prod_{i=1}^n \phi^\xi_{\sigma_i}(l,\bm{u})$ is independent of $\sigma$, we can further simplify this expression to 
\begin{equation}
    E^{(n)}(\bm{u}) = \sum_{l}\sum_{\xi=1}^{N^{(n)}_c} 
    \frac{\lambda_\xi}{n!}  
   \prod_{i=1}^n \phi^\xi_{i}(l,\bm{u}),
\end{equation}
which is Eq.~(6) in the main text. 

\section{Computational cost analysis}
From a computational viewpoint, Eq.~(2) in main text can greatly accelerate calculations of ph-ph interactions. 
The main bottleneck in such calculations is computing the $n$-ph coupling $V^{(n)}(\QQ_1,\QQ_2,...,\QQ_n)$ for many scattering channels starting from the $n$-IFC, $\Phi^{(n)}$. 
When using the uncompressed $n$-ph interactions in Eq.~(1), the computational cost scales as $\mathcal{O} (N_{\Phi} N_{\text{channel}} )$, where $N_{\Phi}$ is the number of non-zero entries in $\Phi^{(n)}$ and $N_{\text{channel}}$ is the number of active scattering channels in $V^{(n)}(\QQ_1,\QQ_2,...,\QQ_n)$. 
In contrast, utilizing the compressed $n$-ph interaction in Eq.~(2), the same calculation has a cost that scales as $\mathcal{O} (N^{(n)}_{c} N_{\text{channel}})$, where $N^{(n)}_{c}$ is the PCP rank. 
Therefore, the computational cost savings from using Eq.~(1) instead of Eq.~(2) is proportional to the compression factor, \mbox{$\gamma^{(n)} = N_{\Phi}/N^{(n)}_{c}$.}
In Fig.~4 of main text, we compare calculations using full (uncompressed) and compressed $n$-ph interactions. The two sets of calculations are identical except for the method used to calculate $V^{(n)}(\QQ_1,\QQ_2,...,\QQ_n)$. 
Therefore, the computational speed-up in Fig. 4 is entirely due to the more efficient computation of $V^{(n)}(\QQ_1,\QQ_2,...,\QQ_n)$ when using compressed $n$-ph interactions. 

\textcolor{black}{
 In thermal conductivity calculations, the speed-up analysis is more complex. 
The total CPU time for computing phonon scattering rates (and thus the thermal conductivity) can be split into two contributions, with respective computational cost:
\begin{align}
&T_1 = \text{cost of evaluating the $n$-phonon interactions }V^{(n)}(\QQ_1,\QQ_2,...,\QQ_n), 
\nonumber
\\
&T_2 = \text{cost of computing the scattering rates from those interaction tensors}.
\nonumber
\end{align}
Our PCP method only accelerates \(T_{1}\), reducing it to $\frac{T_{1}}{k\gamma_{n}}$ ,
where \(\gamma_{n}\) is the compression factor and \(k\) is a small pre-factor (in practice \(k \approx 2\)). The cost \(T_{2}\) is left unchanged. Therefore, the speedup for the phonon scattering rate calculation, defined as the ratio of CPU wall times, is
\[
\mathrm{speedup}
\;=\;
\frac{\,T_{1} + T_{2}\,}{\,T_{1}/(k\,\gamma_{n}) + T_{2}\,}.
\]
In conventional calculations of phonon scattering rates, evaluating \(n\)-phonon interactions is the bottleneck and \(T_{2}\) is negligible. However, since our PCP approach reduces \(T_{1}\) very efficiently, \(T_{2}\) eventually becomes non-negligible. Consequently, there is an upper bound for the speedup:
\[
\mathrm{speedup}
\;<\;
\frac{\,T_{1} + T_{2}\,}{\,T_{2}\,}
\;\approx\;
\frac{T_{1}}{T_{2}}.
\]
For example, in the case of Si, \(T_{1}/(k\,\gamma_{n})\) is smaller than \(T_{2}\), so the observed speedup is dominated by the fixed cost \(T_{2}\) and thus it is close to the upper bound:
\[
\mathrm{speedup}
\;=\;
\frac{\,T_{1} + T_{2}\,}{\,T_{1}/(k\,\gamma_{n}) + T_{2}\,}
\;\approx\;
\frac{\,T_{1} + T_{2}\,}{\,T_{2}\,}
\;\approx\; 260.
\]
For all cubic materials with small unit cells discussed in the manuscript, the wall time of the calculation is dominated by the fixed cost \(T_{2}\), and thus it is close to the upper bound \(T_{1}/T_{2}\).
For ZrO\(_2\), which has 12 atoms in the unit cell, we use a much larger PCP rank (\(N_{c} = 144\)), so that \(T_{1}/(k\,\gamma_{n}) > T_{2}\) and the observed speedup approaches a different limit where the fixed overhead cost \(T_{2}\) becomes irrelevant:
\[
\mathrm{speedup}
\;=\;
\frac{\,T_{1} + T_{2}\,}{\,T_{1}/(k\,\gamma_{n}) + T_{2}\,}
\;\approx\;
k\,\gamma_{3}
\;\approx\; 7200.
\]
This value reflects the true speedup of our PCP approach, which is close to twice the compression ratio: \(k \approx 2\), \(\gamma_{3} \approx 4000\), and thus a speedup close to 8000.
}

\section{DFT calculations and force constants generation}
We generate displacement-force datasets using the {\sc Alamode} package~\cite{alamode}. 
For all materials studied here, we sample 150 random configurations at 300 K using the harmonic phonon dispersion, and calculate forces for these randomly sampled configurations with the PBEsol functional~\cite{pbesol} using VASP~\cite{vasp,vasp-2} with a  2$\times$2$\times$2 $\mathbf{k}$-point grid for all supercells.  
To extract the IFCs, we employ the adaptive LASSO solver implemented in {\sc Alamode}~\cite{alamode}, with hyperparameters optimized by the cross-validation method.
For Si, we use a 5$\times$5$\times$5 supercell containing 250 atoms, with a unit cell lattice constant of \mbox{5.43 \AA} and a kinetic energy cutoff of 500 eV. 
For HgTe, we use a 4$\times$4$\times$4 supercell containing 128 atoms, with a unit cell lattice constant  of \mbox{6.52 \AA} and a kinetic energy cutoff of 300 eV.  
For MgO, we use a 3$\times$3$\times$3 supercell containing 250 atoms, with a unit cell lattice constant of \mbox{4.25 \AA} and a kinetic energy cutoff of 400 eV.  
For TiNiSn, we use a 2$\times$2$\times$2 supercell containing 96 atoms, with a unit cell lattice constant of \mbox{5.87 \AA} and a kinetic energy cutoff of 500 eV.  
For ZrO$_2$, we use a 2$\times$2$\times$2 supercell containing 96 atoms, with a relaxed unit cell lattice constant of $a = \mbox{4.94 \AA},
b = \mbox{5.16 \AA}, c = \mbox{5.08 \AA}$ and a kinetic energy cutoff of 350 eV.

\section{Phonon scattering rate calculations}
We compute the phonon scattering rate as the sum of 3-ph and 4-ph scattering rates \cite{FengTianliLindsay}. 
The 3-ph scattering rate $\tau_{3ph,\QQ}^{-1}$, for a phonon mode $\QQ = \nu \qq$, reads 
\begin{align}
    \tau_{3ph,\QQ}^{-1} &= 
    \frac{\pi \hbar}{4N_{\qq}}\sum_{\QQ_1}\sum_{\QQ_2} 
    \left|V^{(3)}(\QQ_1,\QQ_2,\QQ_3)\right|^2 \frac{1}{\omega_{\QQ}\omega_{\QQ_1}\omega_{\QQ_2}}    \\
    &\times\left[ \frac{1}{2}(1+n_{\QQ_1}+n_{\QQ_2})\delta(\omega_{\QQ}-\omega_{\QQ_1}-\omega_{\QQ_2})+
        (n_{\QQ_1}-n_{\QQ_2})\delta(\omega_{\QQ}+\omega_{\QQ_1}-\omega_{\QQ_2})\right].
        \nonumber
\end{align}
The 4-ph scattering rate $\tau_{4ph,\QQ}^{-1}$, for a phonon mode $\QQ = \nu \qq$, reads
\begin{align}
    \tau_{4ph,\QQ}^{-1} &= 
    \frac{\pi \hbar}{4N_{\qq}}\frac{\hbar}{2N_{\qq}}\sum_{\QQ_1}\sum_{\QQ_2}\sum_{\QQ_3} 
    \left|V^{(4)}(\QQ_1,\QQ_2,\QQ_3,\QQ_4)\right|^2 \frac{1}{\omega_{\QQ}\omega_{\QQ_1}\omega_{\QQ_2}\omega_{\QQ_3}}    \\
    &\times \left[ \frac{1}{6}\frac{n_{\QQ_1}n_{\QQ_2}n_{\QQ_3}}{n_{\QQ}}\delta(\omega_{\QQ}-\omega_{\QQ_1}-\omega_{\QQ_2}-\omega_{\QQ_3})\right .\nonumber\\\nonumber
    &+
    \frac{1}{2}\frac{(1+n_{\QQ_1})n_{\QQ_2}n_{\QQ_3}}{n_{\QQ}}\delta(\omega_{\QQ}+\omega_{\QQ_1}-\omega_{\QQ_2}-\omega_{\QQ_3})   \\
    &+ \left . \frac{1}{2}\frac{(1+n_{\QQ_1})(1+n_{\QQ_2})n_{\QQ_3}}{n_{\QQ}}\delta(\omega_{\QQ}+\omega_{\QQ_1}+\omega_{\QQ_2}-\omega_{\QQ_3}) \right],
        \nonumber
\end{align} 
where $N_{\qq}$ is the number of $\mathbf{q}$-points in the phonon momentum grid, $\omega_{\QQ}$ is the energy of the phonon mode $\QQ$ and $n_{\QQ}$ the corresponding Bose-Einstein thermal occupation. 
The delta function $\delta(x)$ is approximated by a Gaussian function $\e^{-(x/\epsilon)^2 }$, where $\epsilon$ is the Gaussian smearing parameter. 
We set $\epsilon$ = 0.5 meV, 0.2 meV, 0.2 meV, 0.2 meV, 1.0 meV for Si, HgTe, MgO, TiNiSn and ZrO$_2$, respectively.
We employ the random sampling technique in Ref.~\cite{sampling-technique-phscat} to speed up the integration. 
For Si, HgTe, MgO and TiNiSn, we sample 2$\times 10^5$ scattering processes for both 3- and 4-ph scattering rates for each $\tau_{\QQ}^{-1}$, which ensures convergence of the phonon scattering rates. 
For ZrO$_2$, we sample 4$\times 10^4$ scattering processes for 3-ph interactions and 4$\times 10^5$ scattering processes for 4-ph interactions.  

\section{Constrained optimization}
For acoustic phonons, the ph-ph interactions vanish in the long-wavelength limit due to the acoustic sum rule (ASR). However, the thermal occupation of acoustic phonons can be large at finite temperature and diverge in the long-wavelength limit. 
This interplay results in acoustic phonon scattering being non-negligible, even though their interaction matrix elements are vanishingly small. 
We address this problem using a constrained optimization approach inspired by a method we previously developed for compressing $e$-ph interactions using SVD~\cite{compress-eph}.  
The ASR for the 3-IFCs gives the constraint
\begin{align}
\sum_{l_1b_1}\Phi^{\alpha_1,\alpha_2,\alpha_3}_{l_1b_1,l_2b_2,l_3b_3} = 0. 
\label{asr-3ifc}
\end{align}
To enforce the acoustic sum rule~\cite{symmetry-ifc}, we impose $\sum_{l,b} A^\xi_{i}(l,b\alpha) = 0$.
This ASR leads to the following expansion in the long-wavelength limit ($\qq \to 0$),
\begin{align}
\sum_{l_1b_1} e^{i\qq_1 \bm{r}_{l_1}} \Phi^{\alpha_1,\alpha_2,\alpha_3}_{l_1b_1,l_2b_2,l_3b_3} \approx 
i\qq_1 \cdot \sum_{l_1b_1} \Phi^{\alpha_1,\alpha_2,\alpha_3}_{l_1b_1,l_2b_2,l_3b_3}  \bm{r}_{l_1} 
\approx 
i\qq_1  \cdot \bm{F}^{\alpha_1,\alpha_2,\alpha_3}_{l_2b_2,l_3b_3} [\Phi], 
\label{asr-3ifc}
\end{align}
where
\begin{align}
     \bm{F}^{\alpha_1,\alpha_2,\alpha_3}_{l_2b_2,l_3b_3} [\Phi] 
      = \sum_{l_1b_1} \Phi^{\alpha_1,\alpha_2,\alpha_3}_{l_1b_1,l_2b_2,l_3b_3}  \bm{r}_{l_1}
\end{align}
is the ph-ph deformation potential, which is analogous to the $e$-ph deformation potentials discussed in Refs.~\cite{Bardeen1950,Herring1956}. 

To preserve the ph-ph deformation potential $\bm{F}$, we introduce a Lagrange multiplier in the loss function, which becomes:
\begin{align}
        L = \|\tilde{\bm{\Phi}}\left[ \bm{A} \right]-\bm{\Phi}\|^2 + \lambda_{F} 
        \|\bm{F}[\tilde{\bm{\Phi}}\left[ \bm{A} \right]]-\bm{F}[\bm{\Phi}]\|^2. 
\end{align}
In our calculations, we set $\lambda_{F}$ to a large value to numerically enforce the ph-ph deformation potential. 

\mbox{ }
\newpage
\section{Phonon scattering rates for \text{MgO} and \text{TiNiSn}}

\begin{figure}[h]
    \centering
    \includegraphics[width=1.0\linewidth]{ 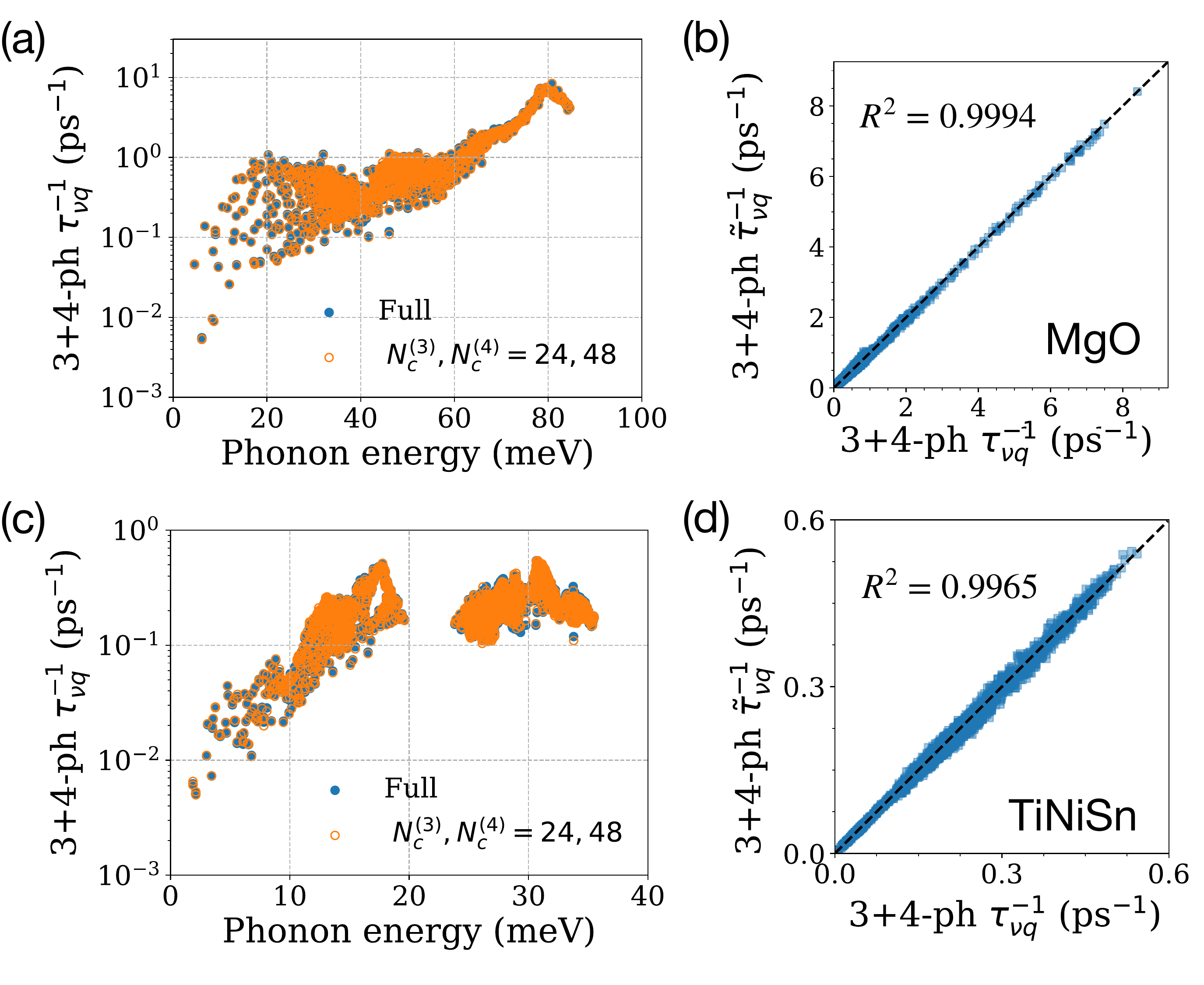}
    \caption{Phonon scattering rates as a function of phonon energy for (a) MgO and (c) TiNiSn. The plots compare results computed with full IFC tensors and compressed IFC tensors (we use $N_c^{(3)}=24$ for MgO and $N_c^{(4)}=48$ for TiNiSn). 
    Corresponding parity plots comparing results from compressed vs. full IFC tensors for 3+4-phonon scattering rates are given the same materials in panels (b) and (d).
    }
    \label{fig:enter-label}
\end{figure}

\mbox{ }
\newpage
\section{Thermal conductivity results}

We calculate the thermal conductivity $\kappa$ at 300 K using momentum grids of $40^3$, $30^3$, $20^3$, $20^3$ and $10^3$ for Si, HgTe, MgO, TiNiSn, and ZrO$_2$ respectively. 
In the following, we show the convergence of $\kappa$  with respect to the PCP decomposition rank for the 3-ph and 4-ph interactions, respectively, in Tables~\ref{tab:1} and \ref{tab:2}. 
For \mbox{$N_c^{(3)}$ = 24} and  \mbox{$N_c^{(4)}$ = 48}, the thermal conductivity predicted using compressed IFC tensors is within 98\% of the value obtained using full 3- and 4-IFC tensors for all materials studied here.

\begin{table}[ht]
\centering
\caption{Calculated thermal conductivity, in units of W/(mK), for Si, HgTe, MgO, TiNiSn. We compare reference results from full (uncompressed) $n$-IFCs (rightmost column) with calculations using compressed 3-ph and 4-ph interactions with different choices of PCP rank for 3-ph interactions. The PCP rank for 4-ph interactions is fixed to $N_c^{(4)}=48$. Thermal conductivities that are within $98\%$ of the uncompressed reference result are shown with bold font.}
\renewcommand{\arraystretch}{1.0}
\setlength{\tabcolsep}{12pt} 
\begin{tabular}{l l l l l l l l  } 
 \hline\hline
   Crystals   &$N_c^{(3)}=4$ \ \  & $N_c^{(3)}=8$  & \ $N_c^{(3)}=16$\ & \ $N_c^{(3)}=24$  & Full $\bm{\Phi}^{(3)}$, $\bm{\Phi}^{(4)}$  \\
 \hline
Si  & 116.6 & \textbf{131.6}  & \textbf{130.7} & \textbf{130.2}  & \textbf{130.9}
\\ 
 \hline
HgTe  & 2.13 & 2.13  & \textbf{2.20} & \textbf{2.22}  & \textbf{2.24}
\\ 
 \hline
MgO  & 59.0 & 44.7  & \textbf{42.8} & \textbf{42.7}  & \textbf{42.3}
\\ 
 \hline
TiNiSn  & 27.7 & 18.7  & \textbf{15.1} & \textbf{15.0}  & \textbf{15.0}
\\ 
\hline
\hline
\end{tabular}
\label{tab:1}
\end{table}

\begin{table}[ht]
\centering
\caption{Calculated thermal conductivity as in Table S1 above but with results given for different choices of PCP rank for 4-ph interactions while fixing the PCP rank for 3-ph interactions to $N_c^{(3)}=24$. Bold font indicates thermal conductivities within $98\%$ of the respective reference result obtained from full $n$-IFC tensors (rightmost column).}
\renewcommand{\arraystretch}{1.0}
\setlength{\tabcolsep}{6pt} 
\begin{tabular}{l l l l l l l l  } 
 \hline\hline
   Crystals   &$N_c^{(4)}=4$ \ \  & $N_c^{(4)}=8$  & \ $N_c^{(4)}=16$\ & \ $N_c^{(4)}=24$ & \ $N_c^{(4)}=32$  & \ $N_c^{(4)}=48$ & Full $\bm{\Phi}^{(3)}$, $\bm{\Phi}^{(4)}$  \\
 \hline
Si  & 123.2 & 132.0  & \textbf{130.7} & \textbf{130.2}  & \textbf{130.3} & \textbf{130.2} & \textbf{130.9}
\\
 \hline
HgTe  & 1.79 & 2.14  & 2.04 & 2.03  & 2.19 & \textbf{2.22} & \textbf{2.24}
\\ 
 \hline
MgO  & 43.8 & \textbf{42.9}  & \textbf{42.8} & \textbf{42.7}  & \textbf{42.7} & \textbf{42.7} & \textbf{42.3}
\\ 
 \hline
TiNiSn  & 15.5 & 15.3 & \textbf{15.0} & \textbf{15.0}  & \textbf{15.0} & \textbf{15.0} & \textbf{15.0}
\\ 
\hline
\hline
\end{tabular}
\label{tab:2}
\end{table}

\newpage

\section{Extrapolation to the thermodynamic limit}
\mbox{}
\begin{figure}[h]
    \centering
    \includegraphics[width=1.0\linewidth]{ 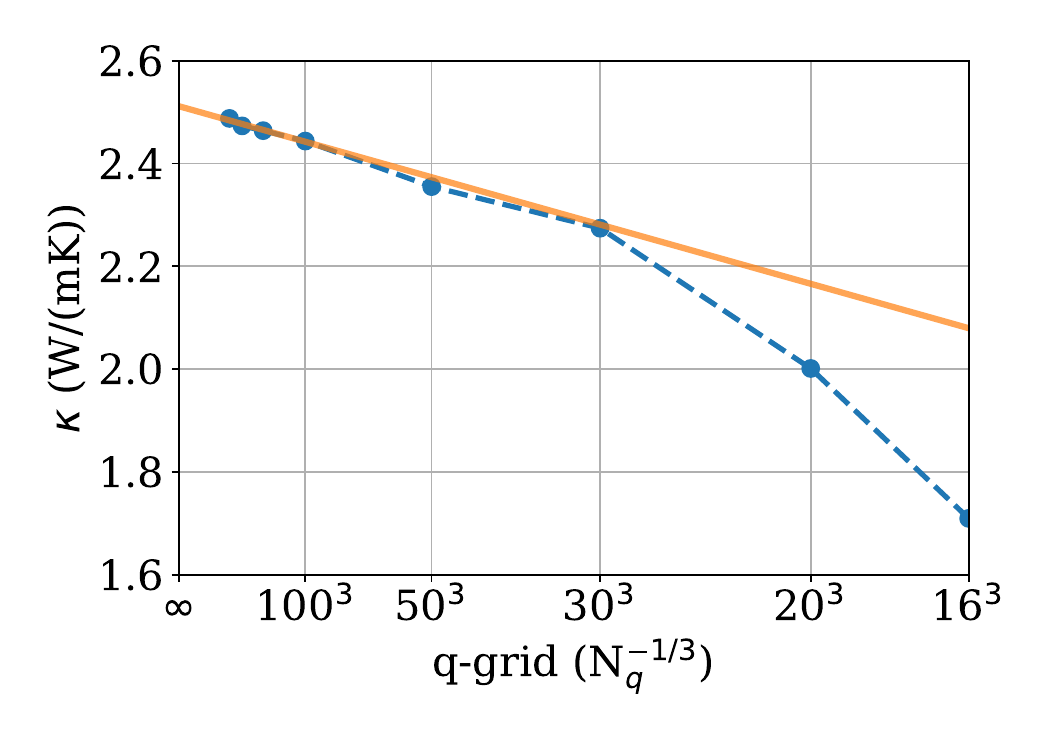}
    \caption{Extrapolation of the thermal conductivity for HgTe at 300 K to the thermodynamic limit (TDL). The largest grid size we computed is $250^3$, which corresponds to the leftmost data point. We extrapolate to the TDL by fitting the thermal conductivity values for the four largest grid sizes using $\kappa(N_q) = \kappa(\infty) - b N_q^{-1/3}$, where $N_q$ is the number of $\mathbf{q}$-points in the grid, and $b$ is the slope. Using this procedure, we obtain an extrapolated value of $\kappa(\infty) = 2.5$ W/(mK), which is 47\% larger than the value using a grid size of $16^3$.}
    \label{fig:enter-label}
\end{figure}
\newpage
\section{Origin of the slow convergence of thermal conductivity for HgTe}
\begin{figure}[h]
    \centering
    \includegraphics[width=1.0\linewidth]{ 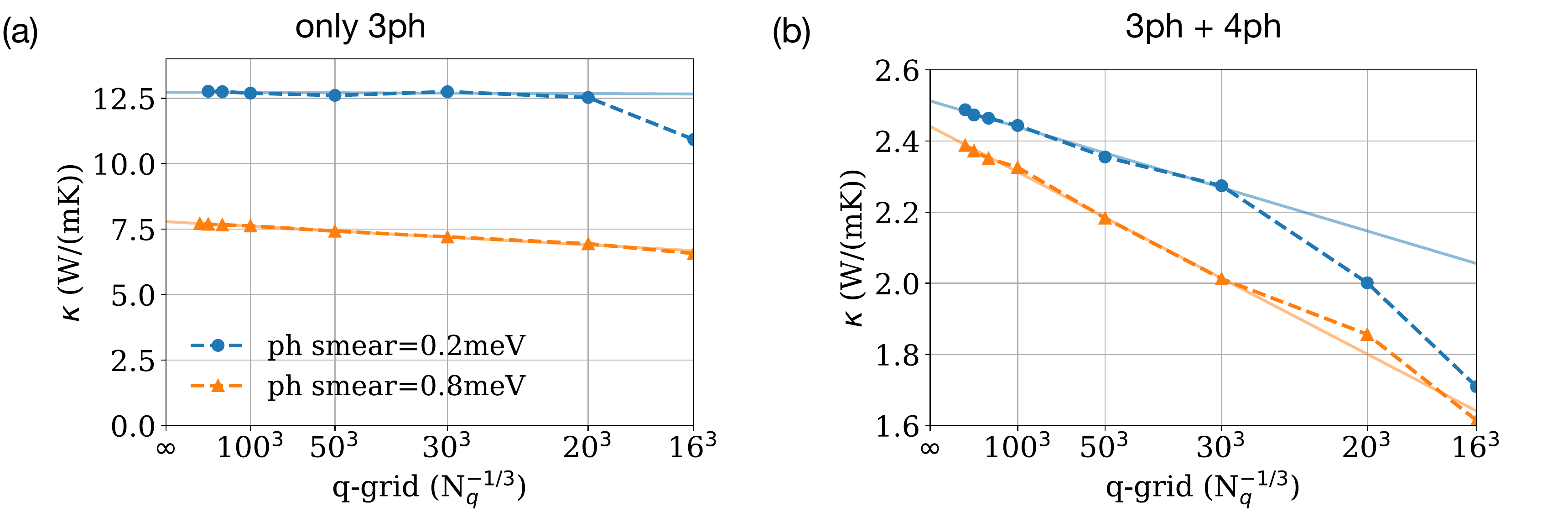}
    \caption{\textcolor{black}{Convergence of the thermal conductivity for HgTe at 300 K with (a) 3-ph and (b) 3+4-ph interactions, using two smearing parameters. Blue and orange markers correspond to smearing values of 0.2 meV and 0.8 meV, respectively.}}
    \label{fig:smearing-34ph-convergence}
\end{figure}

\textcolor{black}{
In HgTe, the 4-ph interactions give an important contribution. Without 4-ph scattering, the thermal conductivity is overestimated by around 400\%, as has been reported previously~\cite{thermal-trans-2}.  
In figure.~\ref {fig:smearing-34ph-convergence}, we calculate the thermal conductivity as a function of grid size for two values of the smearing parameter, 0.2 meV and 0.8 meV, with and without 4-ph interaction. 
We find that the thermal conductivity without 4-ph interactions is greatly overestimated, as is seen by comparing the y-axis values in the two figure panels. 
Figure.~\ref{fig:smearing-34ph-convergence}(a) shows the calculation without 4-ph interactions, where the thermal conductivity is nearly converged for a grid of $20^3$ and is close to the results extrapolated to the thermodynamic limit. 
This confirms that the 3-ph interactions converge rapidly with the grid size.
In Fig.~\ref{fig:smearing-34ph-convergence}(b), in the calculation with 4-ph interactions, the convergence with grid size is slow and persists up to much larger smearing values of 0.8 meV. 
Therefore, we attribute the slow convergence of the thermal conductivity in HgTe to the importance of the 4-ph processes, which are expected to be less smooth than the 3-ph scattering processes in the Brillouin zone and thus require denser grids to reach convergence.
}
\newpage

\section{Number of force constants for 3rd order force constants}
\textcolor{black}{
In this section, we compare the compression factor for our PCP technique, $\gamma^{(3)}$ defined in the manuscript, with the size reduction (or “compression”) resulting from symmetry, defined as
\begin{equation}
  \gamma^{(3)}_{\text{sym}} = \frac{\| \Phi^{(3)}_{\text{full}}\|_0}{\| \Phi^{(3)}_{\text{irr}}\|_0}.
\end{equation}
where $\| \bm{x} \|_0$ indicates the  $L_0$ norm, which measures the number of nonzero elements, and $\Phi^{(3)}_{\text{irr}}$ is the irreducible IFC tensor, which accounts for the reduction of independent entries in the IFC tensor due to symmetry.
In Table~(\ref{tab:1}), we compare the PCP and symmetry-derived compression factors for all materials studied in this work, including ZrO$_2$. The size reduction deriving from symmetry is of order 30–300 for all materials, and thus much smaller than the PCP compression factors, which are in the range of 2000–6000. 
In the case of monoclinic ZrO$_2$, the symmetry-derived compression factor is relatively small ($\gamma^{(3)}_{\text{sym}}$ = 36) due to the low symmetry of ZrO$_2$, while the PCP compression is nearly 100 times greater (to be precise, $93  \approx 3358/36$  times greater). This means that our tensor learning approach becomes even more favorable for low-symmetry crystals. 
}
\begin{table}[ht]
\centering
\caption{Summary of the compression factors of PCP methods and symmetrization for all the studied materials.}
\renewcommand{\arraystretch}{1.0}
\setlength{\tabcolsep}{12pt} 
\begin{tabular}{l l l l l l l l  } 
 \hline\hline
   Crystals   &bravis lattice   & $ \| \Phi^{(3)}_{\text{irr}} \|_0 $ &   $ \| \Phi^{(3)}_{\text{full}} \|_0 $ &  PCP rank $N_c^{(3)}$ &  $\gamma^{(3)}$ & $\gamma^{(3)}_{\text{sym}}$ \\
 \hline
Si  & Cubic & 174  & 54480 & 24  & 2270 & 313
\\ 
 \hline
HgTe  & Cubic & 383  & 54480 & 24  & 2270 & 142
\\ 
 \hline
MgO  & Cubic & 219 & 54480 & 24  & 2270 & 249
\\ 
 \hline
TiNiSn  & Cubic & 515 & 139230 & 24  & 5801 & 270
\\ 
 \hline
ZrO$_2$  & Monoclinic  & 13354 & 483600  & 144 & \textbf{3358} & \textbf{36}
\\ 
\hline
\hline
\end{tabular}
\label{tab:1}
\end{table}

\section{Symmetry loss from compression}

\begin{figure}[h]
    \centering
    \includegraphics[width=1.0\linewidth]{ 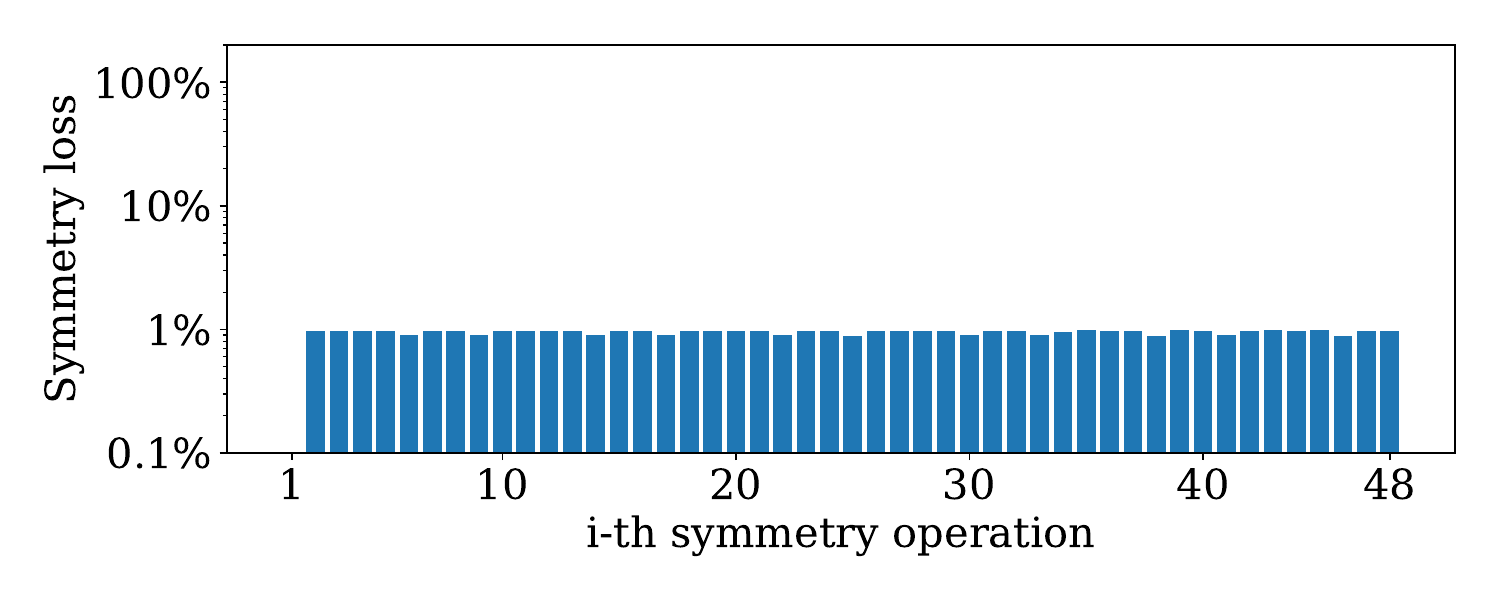}
    \caption{Relative symmetry loss for the compressed 3-ph interactions in Si, using a compression factor $\gamma^{(3)} = 2200$ ($N^{(3)}_c = 24$), given for each symmetry operations in the point group. In all cases, the symmetry loss, $\varepsilon_i$ defined above, is less than 1\%. The first symmetry operation is the identity, which is always satisfied exactly. We use {\sc spglib}~\cite{spglib} to find all the point-group symmetry operations.}
    \label{fig:symmetry}
\end{figure}

We quantify the symmetry loss resulting from compression for the $i$-th symmetry operation,  $\hat{S}_i$, using 
\begin{equation}
    \epsilon_{i} = \frac{\sqrt{\sum_{\QQ_1,\QQ_2,\QQ_3} 
    \left|\tilde{V}^{(3)}(\QQ_1,\QQ_2,\QQ_3) - \hat{S}_i[\tilde{V}^{(3)}](\QQ_1,\QQ_2,\QQ_3)\right|^2}}{\sqrt{\sum_{\QQ_1,\QQ_2,\QQ_3} 
    \left|\tilde{V}^{(3)}(\QQ_1,\QQ_2,\QQ_3) \right|^2} }, 
\end{equation}
where $\hat{S}_i[\tilde{V}^{(3)}]$ is the compressed 3-ph interaction after applying the symmetry operation $\hat{S}_i$. The symmetry transformation follows Ref.~\cite{RevModPhys.40.1}. 
In Fig.~\ref{fig:symmetry}, we show $\epsilon_{i}$ for each symmetry operation in the point group of Si. In all cases, the symmetry loss resulting from compression is less than 1\%. 

